\documentclass{article}
\begin{document}
\title{\bf  Thermodynamics of boson systems related to Dunkl differential-difference operators}
\author{Marcelo R. Ubriaco\thanks{Electronic address:ubriaco@ltp.uprrp.edu}}
\date{Laboratory of Theoretical Physics\\Department of Physics\\University of Puerto Rico\\R\'{\i}o Piedras Campus\\
San Juan\\PR 00931, USA}
\maketitle
\begin{abstract}
We study the thermodynamics of systems based on a Fock space  representation  inspired by the differential-difference 
operators proposed in Ref. \cite{Dunkl}. We calculate  thermodynamic functions as the entropy and heat capacity and compare them 
with the standard boson case. A calculation of the second virial coefficient and the scalar curvature in  two and three dimensions show that
these systems becomes repulsive within an interval of negative values of the reflection operator parameter $\mu_0$. In addition, the stability of 
this system is examined as a function of $\mu_0$.
\end{abstract}

\section{Introduction}

Our work uses as a starting point the differential-difference operator 
defined in Ref. \cite{Dunkl} and used to study
the kernel solutions of the corresponding Laplacian which are known as $h$-harmonic functions. This work was then  extended  in Refs.\cite{Dunkl2}. The differential-difference operator is written
\begin{equation}
{\cal D}_x=\frac{\partial}{\partial x}+\frac{\mu_0}{x}(1-R_x),\label{D}
\end{equation}
where $\mu_0$ is a parameter and the operator $R_x$ is a reflection operator which can be formally
written as $R_x=(-1)^{x\partial/\partial x}$. 
More recently, in Refs.\cite{GMLZ}-\cite{GLZ} this operator was used to study the solutions and symmetries of 
a Hamiltonian with an isotropic harmonic potential in two and three dimensions. It was shown that
this isotropic Dunkl oscillator model is superintegrable and allows separation of variables in the 
usual coordinate systems with solutions in terms of Hermite,Laguerre and Jacobi polynomials.

This paper is organized as follows. In Section \ref{M} we make a correspondence between
the coordinate and the differential-difference  operator in Equation \ref{D} with creation and annihilation operators 
which allow us to define the model we wish to study. In Section \ref{Th} we calculate the partition function leading to
the entropy, heat capacity functions and critical temperature and compare them with the standard Bose-Einstein (B-E) case.
In Section \ref{Vc} we calculate the second virial coefficients in two and three dimensions, and in Section \ref{R}
we calculate the thermodynamic curvature which will tell us about the stability and anyonic behavior of the system.
In Section \ref{Conc} we discuss our results.
\section{The model} \label{M}

Our starting point is very simple, we want to study the consequences of proposing a hamiltonian 
in terms of creation an annihilation operators defined from the correspondence between them and
the coordinate and its derivative respectively. Simply, as it is done in the standard case,
 following the correspondence $a^{\dagger}\leftrightarrow x$
and $a\leftrightarrow \partial/\partial x$ motivated by their commutation relations we define
the hamiltonian
\begin{equation}
{\cal H}=\sum_i\epsilon_i\bar{\phi}_i\phi_i,\label{H}
\end{equation}
where here the  correspondence is given by $\bar{\phi}\leftrightarrow x$ and $\phi\leftrightarrow {\cal D}_x$.
The commutation relation between $\phi_j$ and $\bar{\phi}_j$ is simply given by
\begin{equation}
[\phi_i,\bar{\phi}_j]=\delta_{i,j}(1+2\mu_0 R_i),
\end{equation}
Their action on Fock space is 
\begin{eqnarray}
\bar{\phi}|n>&=&\sqrt{n+1}|n+1>,\\
\phi|n>&=&(\sqrt{n}+\frac{\mu_0}{\sqrt{n}}(1-(-1)^n)|n-1>,
\end{eqnarray}
and therefore the number operator ${\cal N}$ 
\begin{equation}
{\cal N}|n>=n|n>,
\end{equation}
if $n$ is even, and
\begin{equation}
{\cal N}|n>=(n+2\mu_0)|n>,
\end{equation}for $n$ odd.

By defining the operators:$$L_1=(1/2)\bar{\phi}\bar{\phi} ,\;\; L_{-1}=(1/2)\phi\phi,\;\;  L_0=(1/4) (\bar{\phi}\phi+
\phi\bar{\phi)}$$ we get a representation of the $su(1,1)$ algebra
\begin{equation}
[L_m,L_n]=(n-m)L_{m+n}.
\end{equation}
The hamiltonian in Equation \ref{H}
 in terms of the usual operators $a^{\dagger}$ and $a$ becomes
\begin{equation} 
{\cal H}=\sum_i \epsilon_i\left(a^{\dagger}_i a_i+\mu_0(1-(-1)^{\hat{N_i}})\right),\label{H1}
\end{equation}
which is clearly hermitian and $\hat{N_i}$ is the usual number operator. 
\section{Thermodynamic functions}\label{Th}
From Equation \ref{H1} the partition function for this system is given by
\begin{equation}
{\cal Z}=\prod_{l=0}\sum_{n_l=0}e^{-\beta n_l\epsilon_l}e^{-\beta\mu_0\epsilon_l(1-(-1)^{n_l})}z^{n_l},
\end{equation}
which is easy to sum leading to
\begin{equation}
{\cal Z}=\prod_{l=0}\frac{1}{1-e^{-2\beta\epsilon_l}z^2}\left(1+e^{-\beta\epsilon_l}e^{-2\mu_0\beta\epsilon_l}z\right).
\end{equation}
As expected, at $\mu_0=0$ we obtain the B-E partition function $Z=\prod_{l=0}\frac{1}{1-e^{-\beta\epsilon_l}z}$.
The average number of particles
\begin{eqnarray}
<N>&=&\frac{1}{\beta}\frac{\partial}{\partial\mu}\ln{\cal Z},\\
&=&\sum_{l=0}\left(\frac{2}{e^{2\beta\epsilon_l}z^2-1}+\frac{1}{e^{\beta\epsilon_l}e^{2\mu_0\beta\epsilon_l}z+1}\right),
\end{eqnarray}
In particular, the zero momentum distribution is $<n_0>=\frac{z}{1-z}$, which is identical to the B-E case.
Replacing, in the thermodynamic limit, the summation by an integral and expanding in powers of $z$ we obtain
\begin{eqnarray}
\ln{\cal Z}&=&-\ln(1-z)+\frac{1}{\lambda^3}g_{5/2}(\mu_0,z),\\
<N>&=&\frac{z}{1-z}+\frac{1}{\lambda^3}g_{3/2}(\mu_0,z),
\end{eqnarray}
where the functions
\begin{equation}
g_{5/2}(\mu_0,z)=\sum_{n=1}\frac{z^{2n}}{2^{3/2}n^{5/2}}+\left(\frac{1}{1+2\mu_0}\right)^{3/2}\sum_{n=1}\frac{(-1)^{n+1}z^n}{n^{5/2}},
\end{equation}
$g_{3/2}(\mu_0,z)=z\frac{\partial}{\partial z}g_{5/2}(\mu_0,z)$ become  the
standard functions $g_{5/2}(z)$ and $g_{3/2}(z)$ for $\mu_0$ respectively$ and \lambda$ is the thermal wavelength.

Figure 1 is a graph of the functions $g_{5/2}(\mu_0,1)$ and $g_{3/2}(\mu_0,1)$ in the interval $0\leq\mu_0\leq 1$
showing that their values within this interval are smaller than the textbook, $\mu_0=0$, functions.  These functions become singular at $\mu_0=-0.5$ and complex for $\mu_0<-0.5$ restricting therefore the range of values to the interval
$\mu_0>-0.5$.  

The critical temperature
\begin{equation}
T_c=\frac{h^2}{2\pi m k}\left(\frac{<N>}{Vg_{3/2}(\mu_0,1)}\right)^{2/3}.
\end{equation}
Since for $\mu_0>0$ the function $g_{3/2}(\mu_0,1)<g_{3/2}(1)$ the critical temperature $T_c$ is higher than
the critical temperature for B-E case $T_c^{BE}$. For $-0.5<\mu_0<0$, 
$g_{3/2}(\mu_0,1)>g_{3/2}(1)$ and therefore  $T_c<T_c^{BE}$ which means that the system is less attractive.
In general
\begin{equation}
\frac{T_c}{T_c^{BE}}=\left(\frac{2.612}{g_{3/2}(\mu_0,1)}\right)^{2/3},
\end{equation}
and in the range $0\leq\mu_0\leq 1$ the quotient $\frac{T_c}{T_c^{BE}}$ takes values in the interval
$1\leq \frac{T_c}{T_c^{BE}}\leq 1.2$.  For values $\mu_0>>0$ the function $g_(3/2)(\mu_0,1)\approx \frac{1}{\sqrt{2}}g_{3/2}(1)$ and the ratio $\frac{T_c}{T_c^{BE}}\approx 1.26$.
\begin{figure}
\input{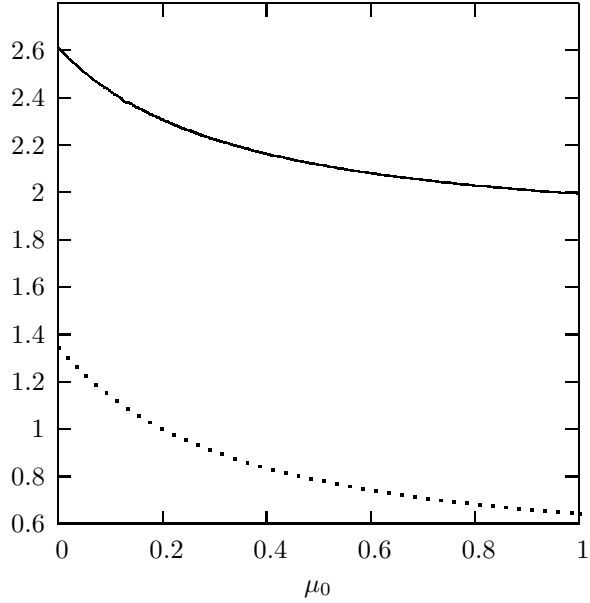}
\caption{The functions $g_{3/2}(\mu_0,1)$ (solid line) and $g_{5/2}(\mu_0,1)$ (dotted line) for $z=1$ and
the parameter $0\leq\mu_0\leq 1$}
\end{figure}
\begin{figure}
\input{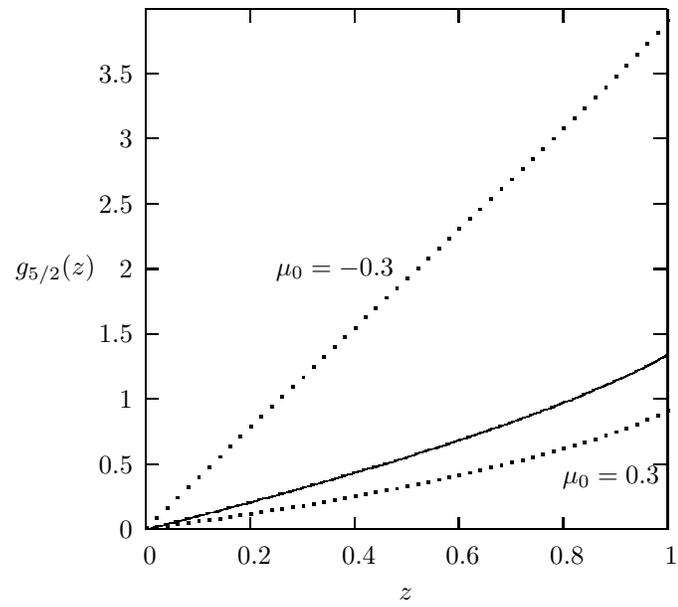}
\caption{The function $g_{5/2}(\mu_0,z)$ for $\mu_0=-0.3$,$\mu_0=0$ (solid line) and $\mu_0=0.3$ in the interval $0\leq z\leq 1$}
\end{figure}
Denoting as $S_+$ and $S_-$ the entropies above and below the critical temperature respectively, we write
\begin{eqnarray}
\lim_{V\rightarrow\infty}\frac{S_+}{V}&=&\frac{5}{2}\frac{k}{\lambda^3}g_{5/2}(\mu_0,z)-\frac{k<N>}{V}\ln z,\\
\lim_{V\rightarrow\infty}\frac{S_-}{V}&=&\frac{5}{2}\frac{k}{\lambda^3}g_{5/2}(\mu_0,1),
\end{eqnarray}

Figure 2 shows a graph of the function $g_{5/2}(\mu_0,z)$ for three values of the parameter $\mu_0$.
For $\mu_0<0$ the function $g_{5/2}(\mu_0,z)>g_{5/2}(z)$ and the entropies $S_+$ and $S_-$  are larger than $S^{BE}$.
In the interval $\mu_0>0$ the entropies $S_{\pm}<S^{BE}$.
Similarly to the standard case the heat capacity above and below the critical temperature is written
\begin{eqnarray}
C_+&=&\frac{15}{4}\frac{k}{\lambda^3}g_{5/2}(\mu_0,z)-\frac{9}{4}\frac{k<N>}{V}\frac{g_{3/2}(\mu_0,z)}{g_{1/2}(\mu_0,z)},\\
C_-&=&\frac{15}{4}\frac{k}{\lambda^3}g_{5/2}(\mu_0,1).
\end{eqnarray}
Since the quotient $\frac{g_{3/2}(\mu_0,z)}{g_{1/2}(\mu_0,z)}$ is, independently of the value of $\mu_0$, almost constant for low $z$ we have that for $\mu_0<0$ ($\mu_0>0$) the heat capacity $C_{\pm}>C^{BE}$ ($C_{\pm}<C^{BE}$).
\section{Virial coefficients} \label{Vc}

In this Section we calculate the second virial coefficients in two and three dimensions. For $D=3$, we expand
\begin{equation}
\ln {\cal Z}=\frac{4\pi V}{h^3}\int_0^{\infty}dp p^{2}\left(-\ln(1-e^{-\beta p^2/m}z^2)+\ln(1+e^{-\beta p^2/2m}e^{-\mu_0\beta p^2/m}z)\right),\label{lnZ}
\end{equation}
 Expanding
the integrand in powers of $z$ and  solving the elementary integrals gives
\begin{equation}
\ln {\cal Z}=\frac{V}{\lambda^3}\left(\frac{z}{(1+2\mu_0)^{3/2}}+\frac{\delta(\mu_0)} {(1+2\mu_0)^3}z^2+...\right),
\end{equation}
where $\delta(\mu_0)=\frac{1}{2^{3/2}}\left((1+2\mu_0)^3-(1/2)(1+2\mu_0)^{3/2}\right)$.
Performing a similar expansion for $<N>$ and after writting the fugacity $z$ in powers of $<N>$ lead to the result
\begin{equation}
pV=kT<N>\left(1-\delta(\mu_0)(\frac{h^2}{2m\pi kT})^{3/2}\frac{<N>}{V}+...\right).
\end{equation}
A similar calculation for $D=2$ gives
\begin{equation}
pA=kT<N>\left(1-\eta(\mu_0)(\frac {h^2}{2m\pi kT})\frac{<N>}{A}+...\right),
\end{equation}
Figure 3 shows a graph of the second virial coefficients for $D=3$ (solid line) and $D=2$ (dotted line) for $-0.5\leq\mu_0\leq 0.1$.
The coefficient $\delta(0)=1/2^{5/2}$, which is the second virial coefficient for the B-E case.
The system behaves as an ideal gas , $\delta(\mu_0)=0$, at the values $\mu_0=-0.5$ and $\mu_0=-0.185$, and between these two values the coefficient $\delta(\mu_0)$ is negative and the system becomes repulsive  reaching the lowest value $\delta(-0.3)=-0.022$. Therefore the interpolation from bosonic to fermionic behavior does not reach the free fermion limit $\delta=-1/2^{5/2}$.  The switch from bosonic to fermionic behavior when the parameter $\mu_0<0$ is consistent with the fact that   the critical temperature is  larger than  $T_c^{BE}$ for
$-0.5<\mu_0<0$. 

 For $D=2$, the ideal gas case is reached at the values $\mu_0=-0.5$ and $\mu_0=-0.25$ and between these two values the virial coefficient $\eta(\mu_0)$ becomes negative but without  reaching the free fermion value $\eta=-1/4$.
For those values $\mu_0>>0$ the virial coefficients functions increase as: $\delta(\mu_0)=2^{3/2}\mu_0^3$ and 
$\eta(\mu_0)=2\mu_0^2$.
 Although this system exhibits anyonic behavior in two and three dimensions, the parameter $\mu_0$ does not interpolate
completely between the free boson and fermion limits.
\begin{figure} 
\input{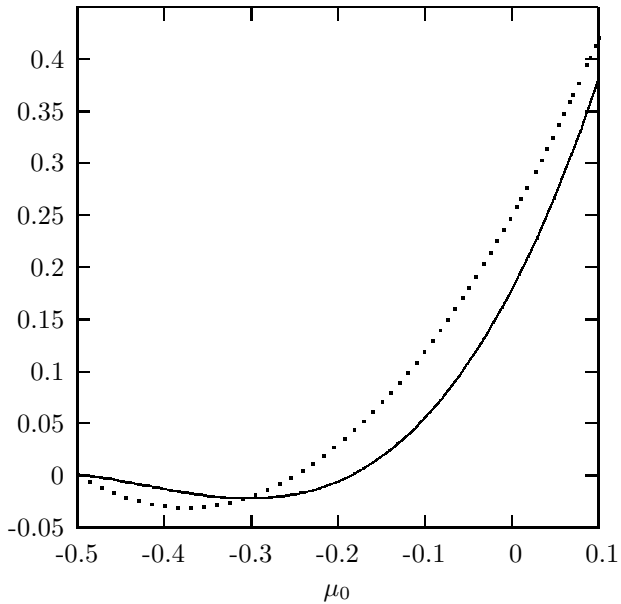}
\caption{The virial coefficients $\delta(\mu_0)$ for $D=3$ (solid line), and $\eta(\mu_0)$ for $D=2$ (dotted line)  as a function of
the parameter $-0.5\leq\mu_0\leq 0.5$}
\end{figure}
\section{Thermodynamic curvature} \label {R}
In this section we calculate the thermodynamic curvature $R$, which is basically the two dimensional curvature in the parameter space spanned by the variables $\beta_1=\beta$ and $\beta_2=-\beta\mu$. The basic geometrical approach to thermodynamics was initiated in Refs.\cite {Ti}-\cite{AN} and extended in \cite{IJKK}-\cite{R} to define a metric and the scalar curvature 
as a measure of the correlations strength of the system. There have been numerous applications of this formalism including
classical and quantum gases \cite{R1}-\cite{BH}, magnetic systems \cite{JM2}-\cite{JJK}, non-extensive statistical mechanics \cite{T-B}-\cite{O}, anyon gas , fractional statistics and deformed boson and fermion systems \cite{MH1},
systems with fractal distribution functions \cite{MRU1}. quantum group invariant systems \cite{MRU2} and systems with $M$-statistics \cite{MRU3}. A calculation of the scalar curvature tell us not only whether the system is attractive (repulsive) from its values $R>0$ ($R<0$) but also about its stability which is obtained from its departure from the classical gas value $R=0$. 
\begin{figure}
\input{figure3}
\caption{The scalar curvature $R$ at $D=3$, in units of $\lambda^3V^{-1}$,  as a function of the fugacity $z$ at constant $\beta$
for the parameter values $\mu_0=-0.3, 0.2$ and the standard case $\mu_0=0$  (solid line).}
\end{figure}
For exponential probability distributions, the metric is simply defined as
\begin{equation}
g_{\alpha\gamma}=\frac{\partial^2\ln Z}{\partial\beta^{\alpha}\partial\beta^{\gamma}},\label{g}
\end{equation}
and the two dimensional scalar  curvature follows from the basic relation
\begin{equation}
R=\frac{2}{det g}R_{1212},
\end{equation}
where $det g=g_{11}g_{22}-g_{12}g_{12}$. Due to the obvious identities$\frac{\partial g_{ij}}{\partial \beta^j}\equiv g_{ij,j}=g_{jj,i}$ the curvature tensor $R_{ijkl}$ reduces to
\begin {equation}
R_{ijkl}=g^{mn}\left(\Gamma_{mil}\Gamma_{njk}-\Gamma_{mik}\Gamma_{njl}\right),
\end{equation}
where the Christoffel symbol $\Gamma_{ijk}=\frac{1}{2}g_{ij,k}$. The curvature $R$ is simply given
by the determinant
\begin{equation} 
R=\frac{1}{2 (detg)^2}\left|\begin{array}{ccc} g_{11} & g_{22} & g_{12}  \\ g_{11,1} & g_{22,1} & g_{21,1} \\
g_{11,2} & g_{22,2}& g_{21,2} \end{array}\right|. \label{Sc}
\end{equation}
The metric components are readily calculated from Equation \ref{lnZ}
\begin{eqnarray}
g_{11}&=& C(\frac{D}{2})(\frac{D}{2}+1)\frac{1}{\beta^{\frac{D}{2}+2}}I_1,\\
g_{12}&=&-C(\frac{D}{2})\frac{1}{\beta^{\frac{D}{2}+1}}I_2,\\
g_{22}&=&\frac{C}{\beta^{D/2}} I_3,
\end{eqnarray}

\begin{figure}
\input{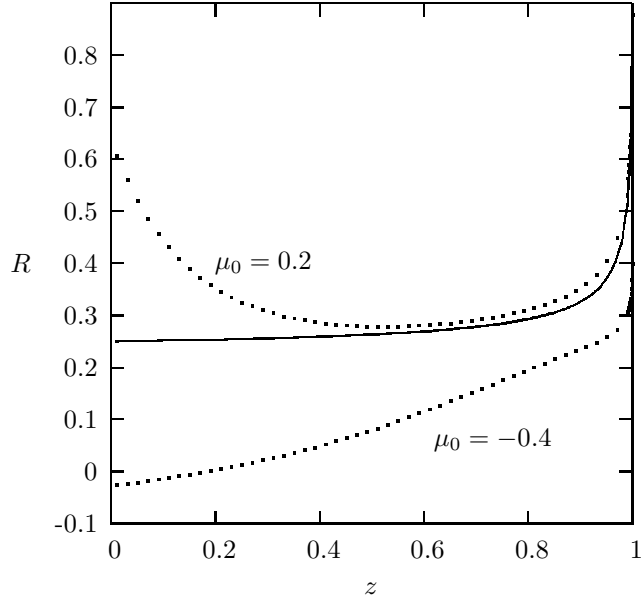}
\caption{The scalar curvature $R$ at $D=2$, in units of $\lambda^2 A^{-1}$,  as a function of the fugacity $z$ at constant $\beta$
for the parameter values $\mu_0=-0.4, 0.2$ and the standard case $\mu_0=0$  (solid line).}
\end{figure}

where the constant $C=\frac{4\pi m A}{h^2}$ for $D=2$, and $C=4\pi V\left(\frac{2m}{h^2}\right)^{3/2}$ for $D=3$. The
 integrals are given by 
\begin{eqnarray}
I_1&=&\int_0^{\infty}dx x^{D-1}\left(-\ln f+\ln g\right),\\
I_2&=&\int_0^{\infty} dx x^{D-1}\left(\frac{2(f-1)}{f}-\frac{g-1}{g}\right),\\
I_3&=&\int_0^{\infty} dx x^{D-1}\left(\frac{4(1-f)}{f}+\frac{4(1-f)^2}{f^2}+\frac{g-1}{g}-\frac{(g-1)^2}{g^2}\right)
\end{eqnarray}
with the functions $f=1-e^{-2x^2}z^2$ and $g=1+e^{-(1+2\mu_0)x^2}z$.
\begin{figure}
\input{figure5}
\caption{The scalar curvature $R$ at $D=3$, in units of $\lambda^3V^{-1}$, as a function of the parameter $\mu_0$
for values of the fugacity $z=0.1, 0.5, 0.9$.}
\end{figure}
From Equation \ref{Sc} we obtain 
\begin{equation}
R=\frac{\lambda^D}{V_D}\frac{\pi^{D/2-1}}{2^{D-1}}(D+2)\frac{I_3I_2^2-2I_1I_3^2+I_1I_2I_4}{\left((D+2)I_1I_3-DI_2^2\right)^2},
\end{equation}
where $V_D$ stands for the area or volume.

Figure 4 shows a graph of the scalar curvature $R$ for the three dimensional case as a function of the fugacity $z$.
For values such that $\mu_0<0$ the system  becomes repulsive at high temperatures ($z\approx 0$) and it is more
stable than the B-E case for all values of $z$. For $\mu_0>0$ the system is always bosonic and it becomes
mora unstable at high temperatures. As expected, there is a singularity in $R$ as $z\rightarrow 1$ at the onset of
Bose-Einstein condensation. For the two dimensional case the behavior of $R$ as a function of $z$, as shown in Figure 5,
is quite similar to the three dimensional case. Figures 6 and 7 are graphs of the scalar curvature $R$ as a function the parameter $\mu_0$  for the values of $z=0.1,0.5,0.9$ for the three and two dimensional cases respectively.  Independently of the value of the fugacity the curvature vanishes at $\mu_0=-0.5$  which is the value that corresponds to the classical,
Maxwell-Boltzmann case.  At high temperatures, there is another value of $\mu_0$ such that the behavior becomes classical, as
for example for $z=0.1$ the curvature $R=0$ at $\mu_0\approx-0.25$, and within the interval $-0.5<\mu_0<0.25$ the curvature becomes negative. At low temperatures the system is  attractive independently of $\mu_0$. Independently of the value of $z$, the instability increases as $\mu_0$ increases.  In general, systems with $\mu_0>0$ are more unstable and therefore more correlated than those with $\mu_0<0$.
\begin{figure}
\input{figure6}
\caption{The scalar curvature $R$ at $D=2$, in units of $\lambda^2A^{-1}$, as a function of the parameter $\mu_0$
for values of the fugacity $z=0.1, 0.5, 0.9$.}
\end{figure}
\section{Conclusions}\label{Conc}
In this manuscript we have proposed a thermodynamic model based on a Fock space defined by making a correspondence with the so called Dunkl differential-difference operators.  The partition function and the occupation number are written in terms of two functions $g_{5/2}(\mu_0,z)$ and $g_{3/2}(\mu_0,z)$ respectively which become the standard $g_{5/2}(z)$
and $g_{3/2}(z)$ as the parameter $\mu_0\rightarrow 0$. These two new functions $g_{\eta}(\mu_0,z)$ impose a lower limit $\mu_0>-0.5$ as a result that they  become complex for $\mu_0<-0.5$. 
A numerical calculation of these functions shows that the critical temperature is higher (lower) than the B-E case $\mu_0=0$ for
the range of values $\mu_0>0$ ($-0.5<\mu_0<0$).
The fact that  the expressions for the entropy and heat capacity are identical than the standard case but  written in terms of
these new functions $g_{\eta}(\mu_0,z)$ help us to conclude that these thermodynamic functions are larger (lower)
than the B-E for $\mu_0<0$ ($\mu_0>0$). A calculation of the virial coefficients for $D=2$ and $D=3$ show
that they vanish at two values of the parameter $\mu_0$ mimicking therefore the behavior of a classical system. Between these two values the virial coefficient becomes negative and therefore the system becomes repulsive but without reaching the
Fermi-Dirac value of $-\frac{1}{2^{5/2}}$ and $-1/4$ for $D=3$ and $D=2$ respectively. These results are consistent with the fact that, for example, the entropy function is larger for $\mu_0<0$ than for $\mu_0\geq 0$.
A calculation of the thermodynamic curvature $R$ gave us a larger picture about the attractive or repulsive behavior
and the stability as a function of either the parameter $\mu_0$ or the fugacity $z$. For all temperature values the curvature graph shows a more (less) correlated system for positive (negative) values of $\mu_0$ as compared with the B-E case. In addition, at low temperatures the behavior is bosonic independently of the value of $\mu_0$ and
it becomes more unstable than the B-E case for all values of $\mu_0>0$. Therefore, by making a correspondence between
Dunkl differential-difference operators with creation and annihilation operators  we have proposed a 
thermodynamic model that exhibits anyonic behavior in two and three dimensions.  Although this model does not
interpolates completely between the B-E and F-D cases it certainly gives a new approach wherein anyonic behavior 
manifest in three dimensions other than previously proposed models based on $M$-statistics \cite{Ng} and quantum group invariance \cite{MRU4}

\end{document}